\documentclass[12pt]{iopart}

\usepackage{graphicx}

\newcommand{\beq}{\begin{equation}}
\newcommand{\eeq}{\end{equation}}

\begin{document}

\title[Precursor diamagnetism in YBa$_2$Cu$_3$O$_{7-\delta}$ with magnetic impurities]{Precursor superconducting diamagnetism in YBa$_2$Cu$_3$O$_{7-\delta}$ with in-plane or intercalated magnetic impurities}

\author{F Soto, N Cot\'on, J D Dancausa, J M Doval, A~Ramos-\'Alvarez, R I Rey, L Rodr\'iguez, C Carballeira, J~Mosqueira, M V Ramallo, F Vidal}

\address{LBTS, Facultade de F\'isica, Universidade de Santiago de Compostela, E-15782 Santiago de Compostela, Spain}

\ead{j.mosqueira@usc.es}

\begin{abstract}
The magnetization around the superconducting transition was measured in YBa$_2$Cu$_3$O$_{7-\delta}$ with magnetic impurities in the CuO$_2$ layers (Cu substituted by Zn or Ni) or between them (Y substituted by Gd or Pr). While some of these impurities have an important effect on the superconducting transition temperature ($T_c$), the precursor diamagnetism observed above $T_c$ is not appreciably affected. This result contrasts with recent observations in a conventional BCS superconductor (La), in which the precursor diamagnetism was found to increase several orders of magnitude with the addition of a small amount (a few atomic percent) of Gd or Pr magnetic impurities.
\end{abstract}

\submitto{\SUST}
\maketitle

\section{Introduction}

The interplay between superconductivity and magnetic order is a long-standing issue which interest has been enhanced by the discovery of the so-called magnetic superconductors, materials where the coupling between both collective states is weak, as it is also the case of the high temperature cuprate superconductors (HTSC).\cite{reviews,recent}  The weak diamagnetism induced by thermal fluctuations above the superconducting transition temperature ($T_c$) is a useful tool to probe some aspects of such interplay, because it does not hide the contribution associated with the magnetic order, as it may be the case below $T_{c}$.\cite{Tinkham} As an example, in Ref.~\cite{EPL06} it was shown that the introduction of magnetic impurities in low-$T_c$ metallic superconductors (LTSC) leads to an increase of the precursor diamagnetism well beyond the uncertainties associated with the normal-state contribution or with possible $T_c$ inhomogeneities. This increase, which was found to be proportional to the impurities concentration, was interpreted in terms of an indirect effect associated to a change in the magnetic coupling between impurities due to the presence of fluctuating Cooper pairs.\cite{EPL06} For LTSC this mechanism was earlier proposed for temperatures below $T_c$.\cite{mecanismo}

Here we probe the effect of magnetic impurities on the fluctuation diamagnetism above $T_c$ in HTSC, an issue that to our knowledge still remains unaddressed. For that, we have studied the prototypical YBa$_2$Cu$_3$O$_{7-\delta}$ (YBCO) with Cu partially replaced by Ni or Zn, or with Y partially replaced by magnetic rare-earths like Gd or Pr. Nonmagnetic Zn$^{2+}$ ($s=0$) substitutes Cu$^{2+}$ ($s=1/2$) in the CuO$_2$ layers,\cite{Yang90} producing a local magnetic moment on the four nearest-neighbors Cu sites arising from the local suppression of short-range antiferromagnetic (AF) correlations.\cite{Zagoulaev,Xiao90}  This magnetic moment is evidenced by the presence of a Curie-like term in the bulk magnetic susceptibility.\cite{Zagoulaev} On the other hand, as determined by x-ray absortion experiments,\cite{Yang90,Bridges90} magnetic Ni substitutes Cu nearly uniformly in both the CuO$_2$ layers and the CuO chains. Both Zn and Ni impurities in the CuO$_{2}$ layers act as strong pairbreakers and are very effective in reducing $T_{c}$.\cite{Maeno87} In contrast, the out-of-plane substitution of Y by a magnetic rare-earth leaves $T_{c}$ almost unaffected\cite{rareearths} except in the case of Pr, which in an amount of $\sim60$\% completely suppresses the superconductivity.\cite{praseodimium} Our measurements then cover in-plane and out-of-plane magnetic substitutions with very different influences on the superconducting transition temperature. Our experimental results show that, in contrast with the LTSC, in YBCO the precursor diamagnetism is not appreciably affected by the presence of magnetic impurities. Although it is beyond the central aim of our work, we will also show that by just taking into account the $T_c$ dependence on the impurities concentration, the observed magnetization roundings above $T_c$ may be explained by conventional mean-field Gaussian-Ginzburg-Landau (GGL) approaches.\cite{LD,Tsuzuki,Yamaji,Lee1,Lee2,Klemm,Ramallo,EPL01}

\section{Experimental details and results}

\begin{figure}[t]
\begin{center}
\includegraphics[scale=.7]{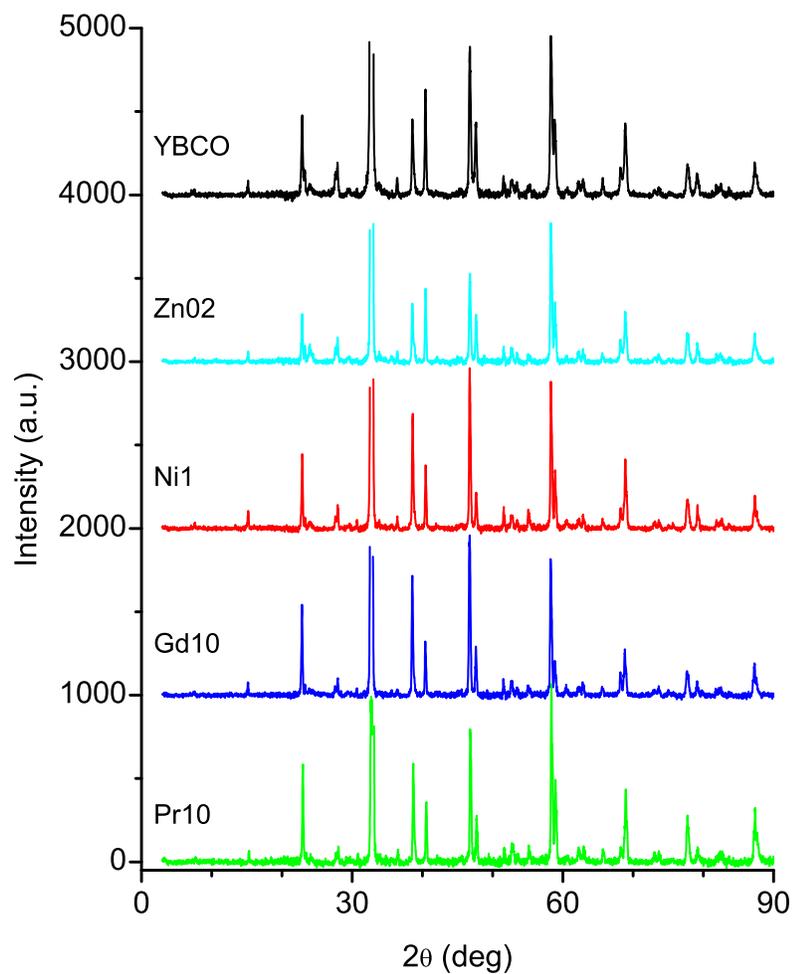}
\caption{X-ray diffraction patterns of some of the samples studied in this work. The observed peak positions are close to the ones of YBCO without impurities, discarding the presence of appreciable impurity phases.}
\label{rx}
\end{center}
\end{figure}

\begin{figure}[t]
\begin{center}
\includegraphics[scale=.7]{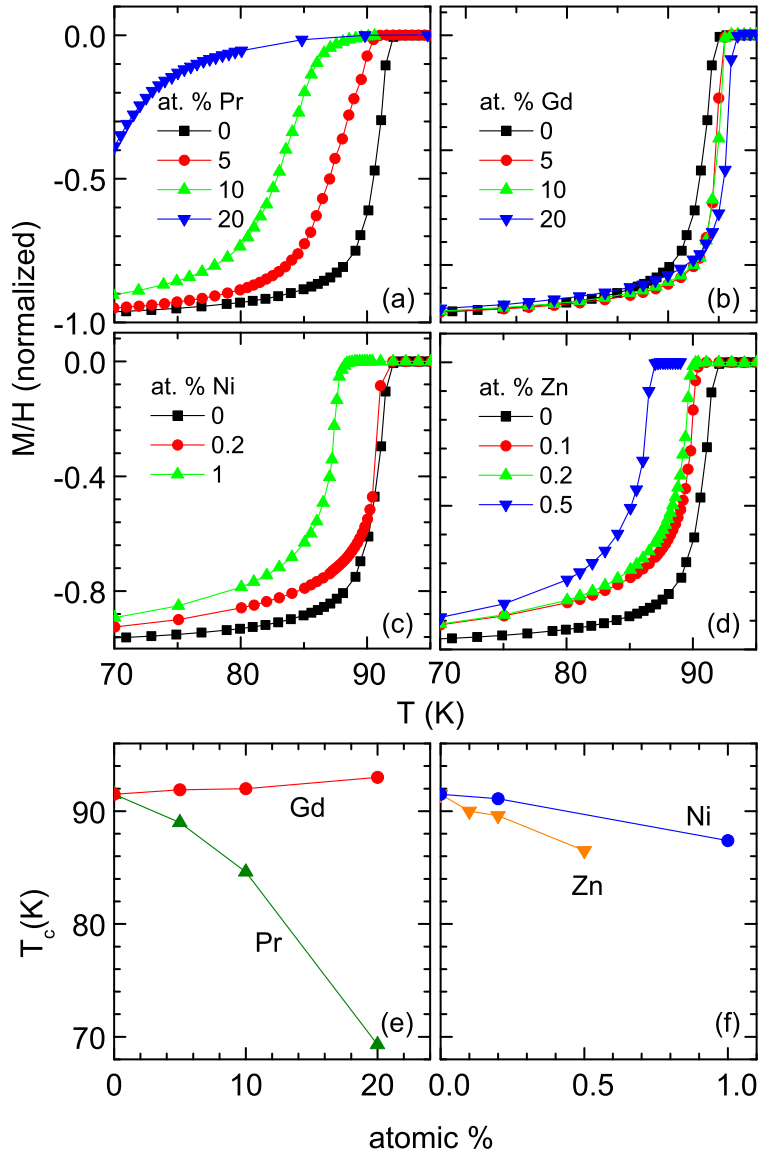}
\caption{(a) to (d) Temperature dependence of the field-cooled magnetic susceptibility of the samples studied in this work, measured with $\mu_0H\stackrel{<}{_\sim}1$~mT. For a better comparison of the transition widths, these data were normalized to the ideal value of -1 at low temperatures. $T_c$ was estimated as the temperature at which the slope is maximum, and the transition width as the difference between $T_c$ and the onset temperature above the transition. The resulting $T_c$ dependences on the doping levels are presented in (e) and (f). The lines are guides for the eyes.}
\label{tcs}
\end{center}
\end{figure}

Powder samples with nominal compositions Y$_{1-x}$Pr$_x$Ba$_2$Cu$_3$O$_{7-\delta}$ ($x\leq0.2$), Y$_{1-x}$Gd$_x$Ba$_2$Cu$_3$O$_{7-\delta}$ ($x\leq0.2$), YBa$_2$(Cu$_{1-x}$Zn$_x$)$_3$O$_{7-\delta}$ ($x\leq0.005$), and YBa$_2$(Cu$_{1-x}$Ni$_x$)$_3$O$_{7-\delta}$ ($x<0.01$) were prepared by the usual solid-state reaction method. To optimize the distribution of dopants, the samples were ground and reacted again several times. Finally, they were held at $\sim425^\circ$C in flowing oxygen several days to attain an oxygen doping level close to the optimal one ($\delta\stackrel{<}{_\sim}0.1$). 
Powder x-ray diffraction (performed with a Rigaku Miniflex II diffractometer) confirmed the orthorhombic structure of YBCO and discarded the presence of appreciable impurity phases (see Fig.~\ref{rx}). As a check of the samples quality we performed measurements of the temperature dependence of the field-cooled (FC) magnetic susceptibility under low applied magnetic fields ($\stackrel{<}{_\sim}1$ mT). These measurements, shown in Fig.~\ref{tcs}, and the subsequent ones of fluctuation effects above $T_c$, were performed with a commercial (Quantum Design) SQUID magnetometer. The $T_c$ value for each sample (shown in Table I) was estimated as the temperature at which the slope of the $M/H$ vs. $T$ curves is maximum. The $T_c$ dependence on the doping levels (presented in Figs.~\ref{tcs}(e,f)) is consistent with previous experiments,\cite{Yang90,Zagoulaev,rareearths,praseodimium} and the corresponding $[T_c(x)-T_c(0)]/x$ range from about $+$5~K for Gd-doping to about $-10^3$~K for Zn-doping. The transition widths, $\Delta T_c$, estimated as the difference between the onset of the low-field diamagnetic transition and $T_c$, are also presented in Table I. From these values, the reduced temperatures, $\varepsilon\equiv\ln(T/T_c)$, below which $T_c$ inhomogeneities may affect the measurements of fluctuation effects, were estimated as $\varepsilon_{\rm inh}=\Delta T_c/T_c$ and compiled in Table I. As it may be seen, with the exception of the most Pr-doped samples, the superconducting transitions are sharp enough to study the fluctuation effects down to the Levanyuk-Ginzburg reduced temperature for the onset of \textit{full-critical fluctuations}, which for YBCO is close to $\varepsilon_{LG}\sim2\times10^{-2}$.\cite{ramallo_elg} This will allow us to investigate the precursor diamagnetism in all the Gaussian region above $T_c$. 

\begin{figure}[t]
\begin{center}
\includegraphics[scale=.7]{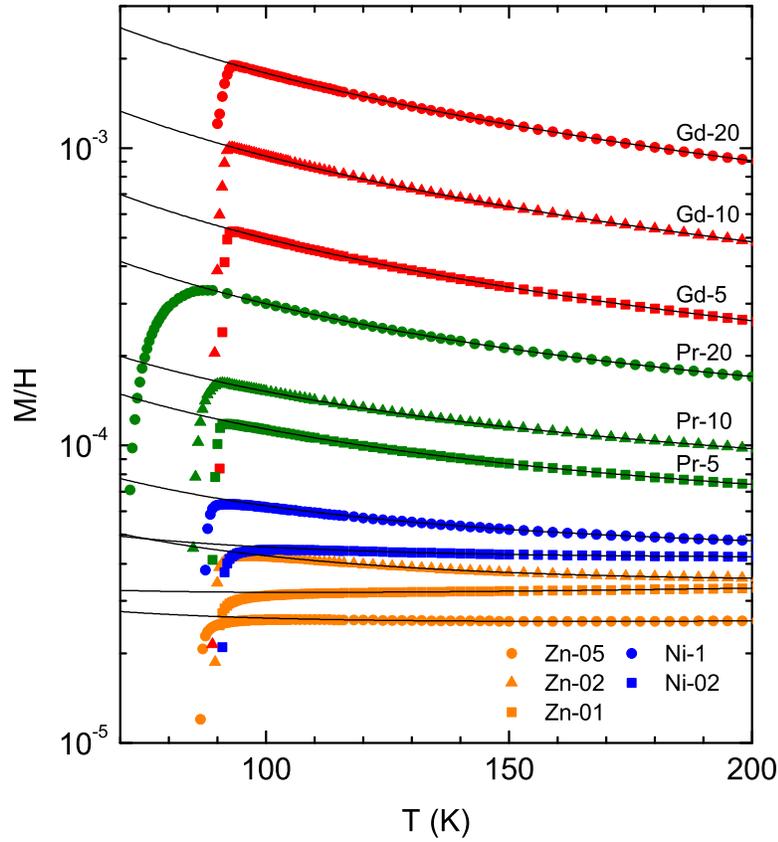}
\caption{Temperature dependence of the magnetic susceptibility of the samples studied in this work up to $\sim 2T_c$, obtained with a 0.5~T applied magnetic field. The logarithmic scale is motivated by the large differences in the normal-state susceptibilities. The lines are the background contributions, obtained by fitting a Curie-like function above $\sim1.5T_c$.}
\label{backs}
\end{center}
\end{figure}

\begin{figure}[t]
\begin{center}
\includegraphics[scale=.7]{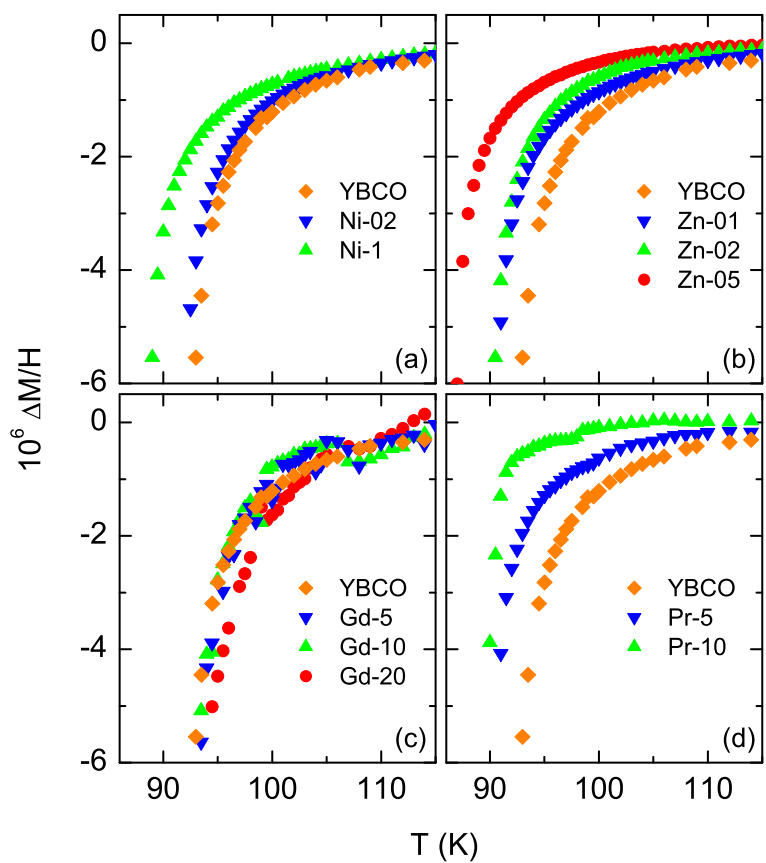}
\caption{Temperature dependence of the magnetic susceptibility above $T_c$ after the normal-state background subtraction.}
\label{delta}
\end{center}
\end{figure}

The magnetization above $T_c$ was measured as a function of temperature by using a 0.5 T magnetic field (Fig.~\ref{backs}). This field is much smaller than the upper critical field for $H\perp ab$ extrapolated to $T=0$~K [$\mu_0H_{c2}(0)\approx270$~T, see below]. Then, the GGL approach for layered superconductors summarized in Sec.~III [valid for $H\ll\varepsilon H_{c2}(0)$] will be applicable in all the $\varepsilon$-range above $\varepsilon_{LG}$. The fluctuation contribution to the magnetization, $\Delta M$, was obtained by subtracting from the as-measured $M(T)$ curves the corresponding normal-state or background contributions, $M_B(T)$. These last were determined by fitting a Curie-like function [$M_B(T ) = a + bT + c/T$, where $a$, $b$ and $c$ are fitting parameters] in a temperature region above $\sim1.5T_c$ (typically above 150 K), where fluctuation effects are expected to be negligible.\cite{EPL02} The background contributions are represented as solid lines in Fig.~\ref{backs}. The resulting fluctuation-induced magnetic susceptibilities are represented in Fig.~\ref{delta}. In view of this figure one may already conclude that the strong increase of the fluctuation diamagnetism observed in LTSC with a similar concentration of magnetic impurities is not present in YBCO. In the next Section we will analyze up to what extent the observed differences may be attributed to the dependence with the impurities concentration of the relevant superconducting parameters.

\section{Comparison with GGL approaches for the fluctuation magnetization in layered superconductors}

Our data will be analyzed in the framework of the GGL-Lawrence-Doniach (GGL-LD) model for Josephson-coupled layered superconductors presented in Ref.~\cite{EPL01}, which includes an \textit{energy cutoff} in the fluctuation spectrum. A justification for the introduction of this cutoff is presented in Ref.~\cite{EPL02}. According to this approach, the fluctuation magnetization for $H\perp ab$ in the weak magnetic field limit [for field amplitudes much smaller than the corresponding $H_{c2}(0)$] is given by\cite{EPL01}
\begin{equation}
\Delta M_\perp=-\frac{\pi k_BT\mu_0H\xi_{ab}^2(0)}{3\phi_0^2s}\left[\frac{1}{\sqrt{\varepsilon(\varepsilon+B_{\rm LD})}}-\frac{1}{\varepsilon^c}\right].
\label{perpcutof}
\end{equation}
Here $B_{\rm LD}\equiv[2\xi_c(0)/s]^2$, $\xi_{ab}(0)$ and $\xi_c(0)$ are the in-plane and transverse coherence length amplitudes, $s=0.585$~nm is the CuO$_2$ layers effective periodicity length in YBCO compounds, $k_B$ is the Boltzmann constant, $\mu_0$ is the vacuum permeability, $\phi_0$ is the flux quantum, and $\varepsilon^c$ is the cutoff constant, which is expected to be close to 0.5.\cite{EPL01,EPL02} This expression predicts the vanishing of fluctuation effects at a reduced temperature, $\varepsilon_{\rm onset}$, close to $\varepsilon^c$. In the non-cutoff case (i.e., for $\varepsilon^c\to\infty$), Eq.~(\ref{perpcutof}) reduces to the well known expression:\cite{LD,Tsuzuki,Yamaji}
\begin{equation}
\Delta M_\perp=-\frac{\pi k_BT\mu_0H\xi_{ab}^2(0)}{3\phi_0^2s\sqrt{\varepsilon(\varepsilon+B_{\rm LD})}}.
\label{perp}
\end{equation}

\begin{figure}[t]
\begin{center}
\includegraphics[scale=.7]{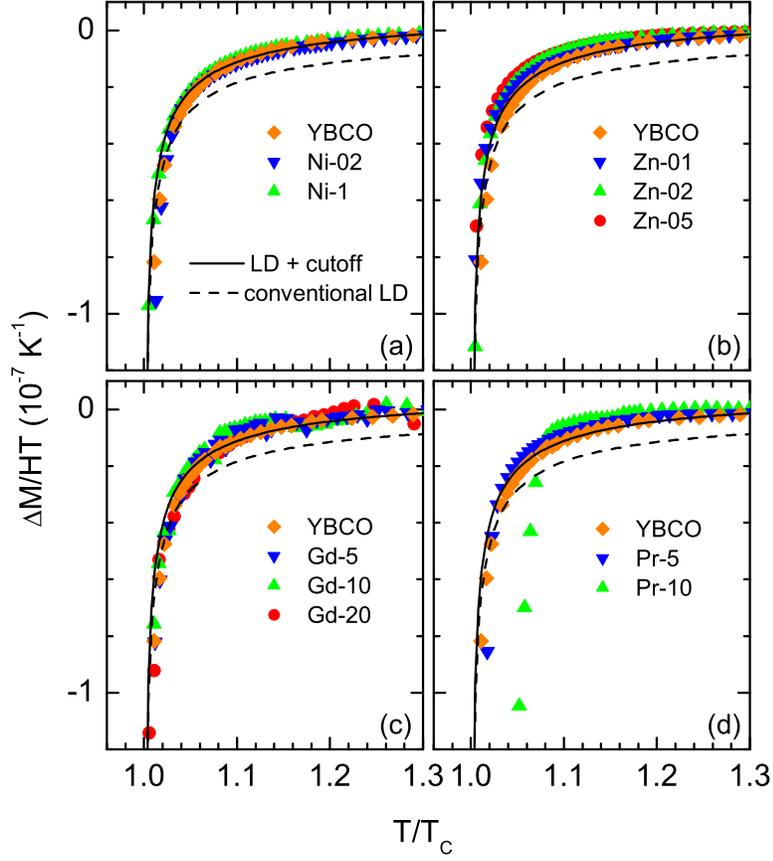}
\caption{$T/T_c$ dependence of the fluctuation magnetic susceptibility (over $T$) of the samples studied in this work. The behavior of the Pr-10 sample may be attributed to its large transition width (see main text for details). The solid lines correspond to the GGL-LD approach with a total-energy cutoff [Eq.~(\ref{poly})] evaluated with typical parameters of the optimally-doped YBCO [$\xi_{ab}(0)=1.1$~nm and $\xi_{c}(0)=0.12$~nm] and a cutoff constant of $\varepsilon^c=0.4$. The conventional GGL-LD approach [Eq.~(\ref{poly}) with $\varepsilon^c\to\infty$] evaluated with the same parameters (dashed lines) does not reproduce the vanishing of fluctuation effects at high reduced temperatures.}
\label{deltanorm}
\end{center}
\end{figure}

To analyze the measurements presented here (obtained in powder samples with the grains randomly oriented) we must perform the angular average of the fluctuation magnetization of an individual crystallite:
\begin{equation}
\Delta M(T,H)=\int_0^{\pi/2}\Delta M(T,H,\theta)\sin\theta\;d\theta.
\label{average}
\end{equation}
Here $\theta$ is the angle between the crystal $c$ axis and the applied magnetic field, and $\Delta M(T,H,\theta)$ may be expressed in terms of the components perpendicular and parallel to the CuO$_2$ layers ($\Delta M_\perp$ and $\Delta M_\parallel$, respectively) as follows:
\begin{equation}
\Delta M(T,H,\theta)=\Delta M_\perp(T,H\cos\theta)\cos\theta+\Delta M_\parallel(T,H\sin\theta)\sin\theta.
\end{equation}
According to the scaling transformation developed in Ref.~\cite{scaling} we may write
\begin{equation}
\Delta M_\parallel(T,H)=\frac{1}{\gamma}\Delta M_\perp\left(T,\frac{H}{\gamma}\right),
\label{para}
\end{equation}
where $\gamma\equiv\xi_{ab}(0)/\xi_c(0)$ is the anisotropy factor. The combination of Eqs.~(\ref{perpcutof}, \ref{average}-\ref{para}) leads to 
\begin{equation}
\frac{\Delta M}{TH}=-\frac{\pi k_B\mu_0\xi_{ab}^2(0)(\gamma^2+2)}{9\phi_0^2s\gamma^2}\left[\frac{1}{\sqrt{\varepsilon(\varepsilon+B_{ \rm LD})}}-\frac{1}{\varepsilon^c}\right].
\label{poly}
\end{equation}
In view of Eq.~(\ref{poly}), the differences in $\Delta M/H$ observed for the samples studied here (Fig.~\ref{delta}) could be attributed to the dependence of $T_c$ or of the coherence lengths on the type and concentration of magnetic impurities. In Refs.~\cite{Pop94, Zhuo97,Tomimoto99} it was shown that the coherence lengths do not change appreciably with respect to pure YBCO, at least for the impurities and concentrations studied. Then, to compensate the differences in $T_{c}$, in Fig.~\ref{deltanorm} we represent $\Delta M/H$ normalized by $T$ against $T/T_c$. As it may be clearly seen, the curves corresponding to different impurities and concentrations scale with the one of pure YBCO.\cite{nota} This contrasts with the above mentioned behavior of conventional BCS superconductors (La doped with Pr), for which similar concentrations of magnetic impurities lead to a strong increase of the precursor diamagnetism.\cite{EPL06} This increase was explained by taking into account that the interaction between the Pr ions is mediated by the electronic sea:\cite{EPL06} According to the RKKY model, this interaction is long-range and dependent on the normal carriers density of states (DOS). The change in DOS induced by fluctuating Cooper pairs alters the interaction between Pr ions and modifies the contribution to the magnetic susceptibility proportionally to the fluctuation superfluid density, $n_s$. As a result, the precursor diamagnetism should include an \textit{indirect} contribution proportional to $n_s$ and to the impurities concentration, which is confirmed by the experiments.\cite{EPL06} Our present results suggest that this mechanism plays a negligible role in the HTSC. In-plane magnetic impurities (Ni or Zn) just interact via exchange with the surrounding Cu ions, and the change in DOS induced by fluctuations seems to leave unchanged their contribution to the normal-state susceptibility. In the case of magnetic impurities away from the CuO$_2$ layers (Gd or Pr), a negligible influence of the carriers sea is even more justified given the weak interlayer coupling in these materials.

For completeness, we compare the data in Fig.~\ref{deltanorm} with the GGL-LD approach for the diamagnetism induced above $T_c$ by thermal fluctuations in layered superconductors. The solid lines correspond to Eq.~(\ref{poly}) and were plotted by using the coherence length amplitudes typical of optimally-doped YBCO [$\xi_{ab}(0)=1.1$~nm, or equivalently $\mu_0H_{c2}(0)=270$~T, and $\xi_c(0)=0.12$~nm],\cite{EPL01,Ando02,Ramshaw12} and a \textit{cutoff constant} $\varepsilon^c=0.4$. As it may be seen, the agreement is excellent in the whole temperature range above $T_c$. Moreover, the cutoff constant (which is close to the $\varepsilon$-value for the onset of fluctuation effects, $\varepsilon_{\rm onset}$) is within the ones found in the fluctuation diamagnetism of other superconducting families, including highly anisotropic HTSC,\cite{HTSC,HTSC2} low-$T_c$ metallic superconductors,\cite{PbIn} MgB$_2$,\cite{MgB2} NbSe$_2$,\cite{NbSe2} and the recently discovered iron-based superconductors.\cite{pnictide}  It is also close to the estimation presented in Ref.~\cite{EPL02} ($\varepsilon^c=0.55$). Our results confirm the proposal in this last work that the onset temperature for the fluctuation effects, $T_{\rm onset}\approx T_c{\rm exp}(\varepsilon^c)\approx1.5T_c$, arises in the limits encountered at high-$\varepsilon$ to the shrinkage of the superconducting wave function to lengths of the order of the pairs size, $\xi_0$.\cite{nano} Moreover, our results are also consistent with the observation that in optimally-doped YBCO and NdBa$_2$Cu$_3$O$_7$ the onset temperature for the Nernst effect follows the change in $T_c$ upon Zn- and Ni- doping, respectively.\cite{Xu05,Johannsen07} They also support the recent proposal that the fluctuation Nernst effect in Ca- and Zn-doped YBCO may be explained by GGL descriptions.\cite{Kokanovic09} Finally, note that disorder induced by electron irradiation in optimally and underdoped YBCO leads to a significant decrease of $T_c$ (together with an increase of $\Delta T_c/T_c$) but leaves the onset temperature of the Nernst signal almost unchanged.\cite{Rullier06} This contrasts with the effect associated to magnetic impurities observed here and in Refs.~\cite{Xu05,Johannsen07}, which could suggest that the disorder induced by irradiation is different in nature.

\section{Conclusions}

We have presented measurements of the magnetization just above $T_c$ in YBCO with magnetic impurities within the CuO$_2$ layers (Zn or Ni replacing Cu), or between them (Pr or Gd replacing Y). The rounding associated with thermal fluctuations seems to be unaffected by the presence of these impurities, regardless of their type and placement. This contrasts with the behavior of some conventional BCS superconductors, in which magnetic impurities induce a strong increase of the precursor diamagnetism.\cite{EPL06} The mechanism proposed in Ref.~\cite{EPL06} to explain such behavior (i.e., a  change in the magnetic coupling between impurities due to the presence of fluctuating Cooper pairs) seems to play a negligible role in the cuprates. Our results further confirm the applicability of mean-field GGL descriptions for the effect of thermal fluctuations above the superconducting transition in HTSC.

\ack

This work was supported by the Spain's MICINN (Project No.~FIS2010-19807), and by the Xunta de Galicia (Project Nos.~2010/XA043 and 10TMT206012PR). N. C. and A. R-A. acknowledge financial support from Spain's MICINN trough a FPI grant.

\newpage

\begin{table}[b]
\begin{center}
\begin{tabular}{ccccc}
\hline
Sample  & Composition & $T_c$ & $\Delta T_c$ & $\varepsilon_{\rm inh}$  \\
 & & (K) & (K) &  $(\times10^{-2}$) \\
\hline
YBCO & YBa$_2$Cu$_3$O$_{7-\delta}$ & 91.5 & 0.6  & 0.7 \\
Pr-5 & Y$_{0.95}$Pr$_{0.05}$Ba$_2$Cu$_3$O$_{7-\delta}$ & 88.9 & 2.5  &  2.8 \\
Pr-10 & Y$_{0.9}$Pr$_{0.1}$Ba$_2$Cu$_3$O$_{7-\delta}$ & 84.6 & 6.9   & 8.2 \\
Pr-20 & Y$_{0.8}$Pr$_{0.2}$Ba$_2$Cu$_3$O$_{7-\delta}$ & 69.3 & 20.5   & 29.6\\
Gd-5 & Y$_{0.95}$Gd$_{0.05}$Ba$_2$Cu$_3$O$_{7-\delta}$ & 91.9 & 0.5  &  0.5\\
Gd-10 & Y$_{0.9}$Gd$_{0.1}$Ba$_2$Cu$_3$O$_{7-\delta}$ & 92.0 & 0.5  &  0.5\\
Gd-20 & Y$_{0.8}$Gd$_{0.2}$Ba$_2$Cu$_3$O$_{7-\delta}$ & 93.0 & 0.5  &  0.5\\
Zn-01  & YBa$_2$(Cu$_{0.999}$Zn$_{0.001}$)$_3$O$_{7-\delta}$ & 90.0 & 1.0 &  1.1 \\
Zn-02  & YBa$_2$(Cu$_{0.998}$Zn$_{0.002}$)$_3$O$_{7-\delta}$ & 89.6 & 0.6  &  0.7\\
Zn-05  & YBa$_2$(Cu$_{0.995}$Zn$_{0.005}$)$_3$O$_{7-\delta}$ & 86.5 & 0.7  & 0.8\\
Ni-02  & YBa$_2$(Cu$_{0.998}$Ni$_{0.002}$)$_3$O$_{7-\delta}$ & 91.1 & 1.0  &  1.1\\
Ni-1 & YBa$_2$(Cu$_{0.99}$Ni$_{0.01}$)$_3$O$_{7-\delta}$ &  87.4 & 1.6   & 1.8\\
\hline
\end{tabular}
\caption{Critical temperatures and transition widths of the samples studied in this work, as determined from the data in Fig.~\ref{tcs}. $\varepsilon_{\rm inh}$ is the reduced temperature below which $T_c$ inhomogeneities are expected to appreciably affect the rounding associated with fluctuation effects.}
\end{center}
\end{table}

%
%

%
%

%
%

%
%

%
%


\begin{thebibliography}{60}
\bibitem{nota}It is worth noting the marked deviation observed in Fig.~\ref{deltanorm} below $T/T_c\approx1.08$ in the Pr-10 sample. However, as it may be seen in Table I, this $T/T_{c}$ value is close to $1+\varepsilon_{\rm inh}$ and this effect may be attributed to $T_c$ inhomogeneities.
\bibitem{nano}As we have earlier stressed (Vidal F, Ramallo M V, Mosqueira J and Carballeira C 2003 {\it Int. J. Mod. Phys. B} {\bf17} 3470) this is not in contradiction with the fact that superconductivity is not destroyed in particles with diameters smaller than the $\xi_0$ value of the \textit{bulk} material, because in such small superconductors $\xi_0$ may also be strongly reduced with respect to the bulk value.
\end{thebibliography}
\end{document}